\documentclass[pra,aps, twocolumn,showpacs]{revtex4}
\usepackage{amsmath}
\usepackage{amssymb}
\usepackage{epsfig}

\begin{document}

\title{NMR implementation of Factoring Large Numbers with Gau\ss{ }Sums: Suppression of Ghost Factors}
\author{Xinhua Peng}
\author{Dieter Suter}
\email{dieter.suter@uni-dortmund.de}
\affiliation{Fachbereich Physik, Universit\"{a}t Dortmund, 44221 Dortmund, Germany}
\date{\today}

\begin{abstract}
Finding the factors of an integer can be achieved by various experimental techniques,
based on an algorithm developed by Schleich et al., which uses specific properties of  Gau\ss{ }sums. 
Experimental limitations usually require truncation of these series, but if the truncation parameter
is too small, it is no longer possible to distinguish between factors and so-called ``ghost" factors.
Here, we discuss two techniques for distinguishing between true factors and ghost factors
while keeping the number of terms in the sum constant or only slowly increasing.
We experimentally test these modified algorithms in a nuclear spin system, using NMR.
\end{abstract}

\pacs{03.67.Lx}

\maketitle

% \section{Introduction}

\emph{Introduction ---}  Factorization of large numbers is a computationally hard problem:
the computational resources required to accomplish this task increase exponentially 
with the number of digits \cite{Knuth:1998aa} for all algorithms discovered until 1994.
Then Peter Shor developed an algorithm that can solve the task in polynomial time.
This algorithm requires a computational device that operates according to the 
laws of quantum mechanics, storing information in quantum states and performing
logical operations as unitary evolutions under suitable Hamiltonians \cite{shor1994}.
Experimental implementations of Shor's factorization algorithm were demonstrated first 
with nuclear spins as qubits \cite{Vandersypen:2001aa}, 
and recently with photonic qubits \cite{Lu:2007fk,Lanyon:2007lr}.

More recently, another factorization algorithm
was proposed by Schleich and co-workers \cite{MerkelCrasser2006,MAGPS2006, Merkel1,Merkel2}
which uses properties of Gau\ss{ }sums. 
A complete normalized quadratic Gau\ss{ }sum is defined by 
\begin{equation}
\mathcal{A}_N^{l-1}(l) = \frac{1}{l}\sum_{m=0}^{l-1}\exp\bigg[2\pi i m^2\frac{N}{l}\bigg]
\label{Com.Gauss}
\end{equation}
where $N$ is the integer to be factorized and $l$ is the trial factor. 
If $l$ is a factor of $N$, i.e., $N/l$ is an integer, the resulting sum is $\vert  \mathcal{A}_N^{l-1}(l) \vert = 1$. 
In all other cases, $\vert  \mathcal{A}_N^{l-1}(l) \vert < 1$. 

The number of terms that has to be evaluated for the complete Gau\ss{ }sum of Eq. (\ref{Com.Gauss})
grows as $\sum_{l=1}^{\sqrt{N}} l= \frac{1}{2} \sqrt{N} (\sqrt{N} -1) \propto N$.
A factorization algorithm on the basis of Eq. (\ref{Com.Gauss}) is thus computationally very expensive.
However, in most cases, a complete evaluation is not necessary.
Recent experimental implementations using NMR \cite{Mahesh:2007aa,Mehring:2007aa}, 
cold atoms \cite{Gilowski-M:aa} and ultra-short laser pulses \cite{Bigourd-D:2007aa} 
have successfully demonstrated that it is usually possible to truncate the sums after a relatively
small number of terms. We write the corresponding truncated sums as 
\begin{equation}
\mathcal{A}_N^M(l) = \frac{1}{M+1}\sum_{m=0}^M\exp\bigg[2\pi i m^2\frac{N}{l}\bigg] ,
\label{Tru.Gauss}
\end{equation}
with a constant truncation parameter $M$ for each argument $l$, instead of the upper limit $l-1$ in the complete Gau\ss{ }sum of Eq. (\ref{Com.Gauss}). 
Accordingly, only $M \sqrt{N}$ terms have to be added, greatly improves the efficiency and precision of the experiments. 
However, the truncation of the Gau\ss{ }sum weakens the discrimination of the factors from nonfactors, 
resulting in the appearance of ``ghost" factors, whose Gau\ss{ }sums are close to unity. 
The requirement of suppressing these  ``ghost" factors thus sets a lower limit on the choice of the truncation parameter $M$. 

The choice of an optimal truncation parameter $M$ was discussed in a recent paper by
Stefanak et al. \cite{M-Stefanak:370aa}.
They found an upper bound on the truncation parameter $M \sim \sqrt[4]{N}$, which represents a sufficient and necessary condition for the success of the Gau\ss{ }sum factorization scheme. 
Here, we experimentally explore this issue, using liquid-state NMR for the evaluation of the Gau\ss{ }sums.
Furthermore, we demonstrate additional possible ways of reducing the truncation parameter $M$,
while keeping excellent contrast between factors and non-factors.

% \section{Effect of truncation}

\emph{Effect of truncation ---} The choice of the truncation parameter $M$ plays a crucial role in the success of the Gau\ss{ }sum factorization scheme. 
In the previous experiments, the visibility of the resulting factorization interference pattern was high enough for successful factorization of, e.g., the eight-digit number $N = $ 52882363 
by logarithmically choosing the truncation parameter $M = \ln N$ \cite{Mehring:2007aa,Mahesh:2007aa}. 
However, in some cases, we also observed some``ghost" factors. 
Like in Ref.\cite{M-Stefanak:370aa}, we define those trial factors for which the absolute value of
the truncated Gau\ss{ }sum is larger than the threshold value $1/ \sqrt{2}$
as  ``ghost" factors. 

\begin{figure}[tbh]
\begin{centering}
\includegraphics[width=0.99\columnwidth]{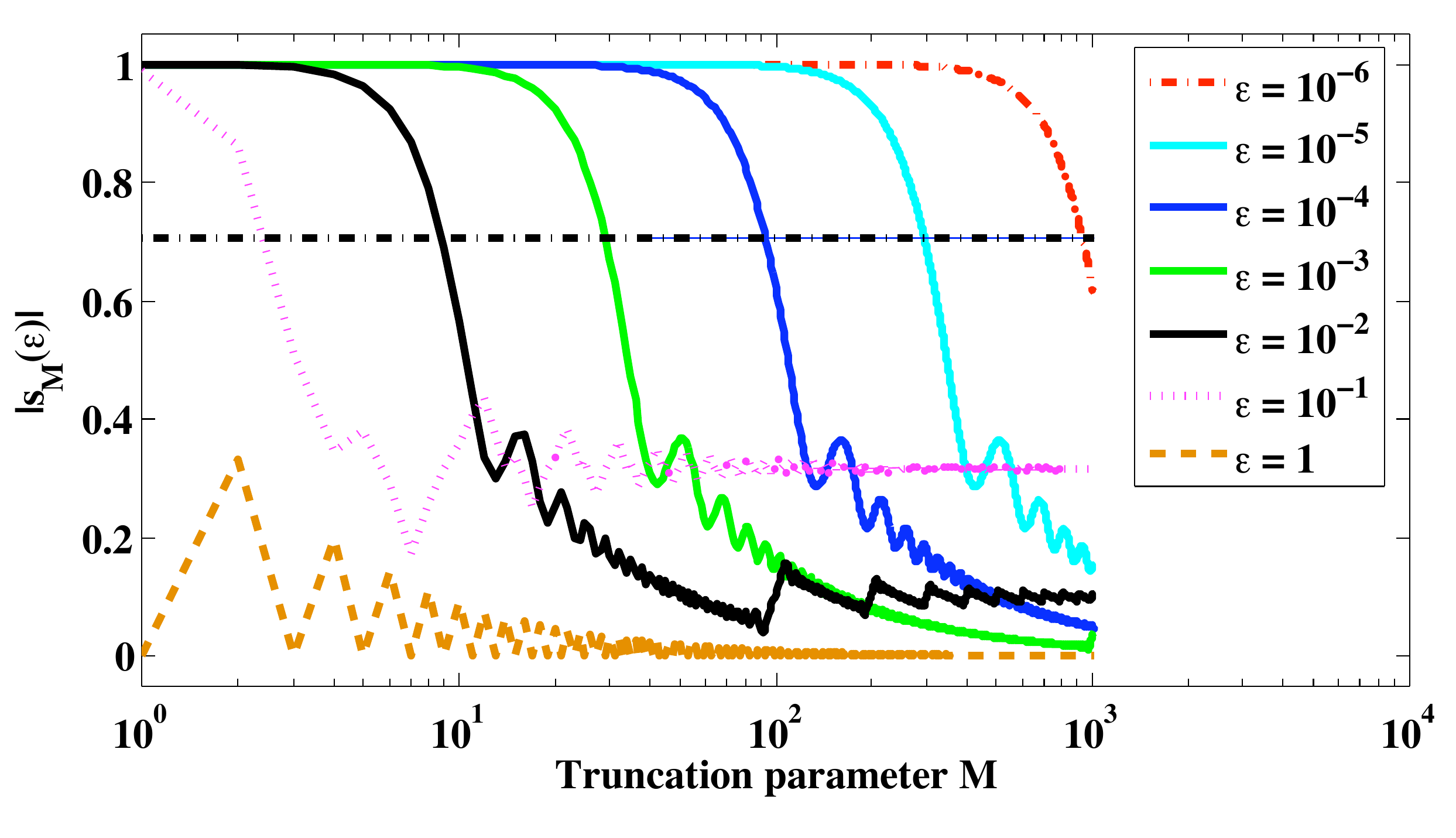}
\end{centering}
\caption{(color online).  Absolute values $|s_M(\epsilon)|$ of the normalized curlicue function 
vs. the truncation parameter $M$ for different values of $\epsilon$.} 
\label{svsM}
\end{figure}

As Stefanak et al. showed \cite{M-Stefanak:370aa}, this sum behavior for different trial factors is best analyzed by considering the fractional part of $2N/l$,
\begin{equation}
\epsilon (N, l) = \frac{2N}{l} - 2k
\end{equation}
with $| \epsilon | \le 1$. 
Here $2k$ is the closest even integer to $2N/l$. 
Since $\exp (i2\pi m^2 k) =1$, the Gau\ss{ }sum (\ref{Tru.Gauss}) can be rewritten as 
\begin{equation}
\mathcal{A}_N^M (l) = s_M(\epsilon) \equiv \frac{1}{M+1} \sum_{m=0}^M \exp (i\pi m^2 \epsilon) ,
\label{s_M}
\end{equation}
where $s_M(\epsilon)$  is the normalized curlicue function,
which has the property:
\begin{eqnarray}
s_M(\epsilon) & = &  \left\{ \begin{array}{lll}
 1,& \epsilon = 0 & \textrm{ for all factors},\\
 \frac{1}{\sqrt{2}}, & \epsilon = 0.5 & \textrm{ for threshold non-factors}, \\
 0, & \epsilon \rightarrow 1 & \textrm{ for typical non-factors}.
\end{array} \right. 
\label{s_M2}
\end{eqnarray}
Here three different classes for the trial factors are defined \cite{M-Stefanak:370aa}. For the class of  the ghost factors, the curlicue function depends on the truncation parameter $M$: 
\begin{eqnarray}
s_M(\epsilon) & \overset{\epsilon \rightarrow 0}{ \longrightarrow} &  \left\{ \begin{array}{ll}
 1,&  \textrm{for a small $M$},\\
 0, & \textrm{for a very large $M$}. 
\end{array} \right. 
\label{s_M3}
\end{eqnarray}
Ghost factors occur when $\epsilon$ is very close to zero and the Gau\ss{ }sum is truncated 
after too few terms. 
Figure \ref{svsM} illustrates this behavior: 
for a given $\epsilon$, the number of terms $M$ needed to suppress the value of $s_M(\epsilon)$
below the threshold $1/\sqrt{2}$ is $\approx 1/\sqrt{\epsilon}$.

\begin{figure}[tbh]
\begin{centering}
\includegraphics[width=0.9\columnwidth]{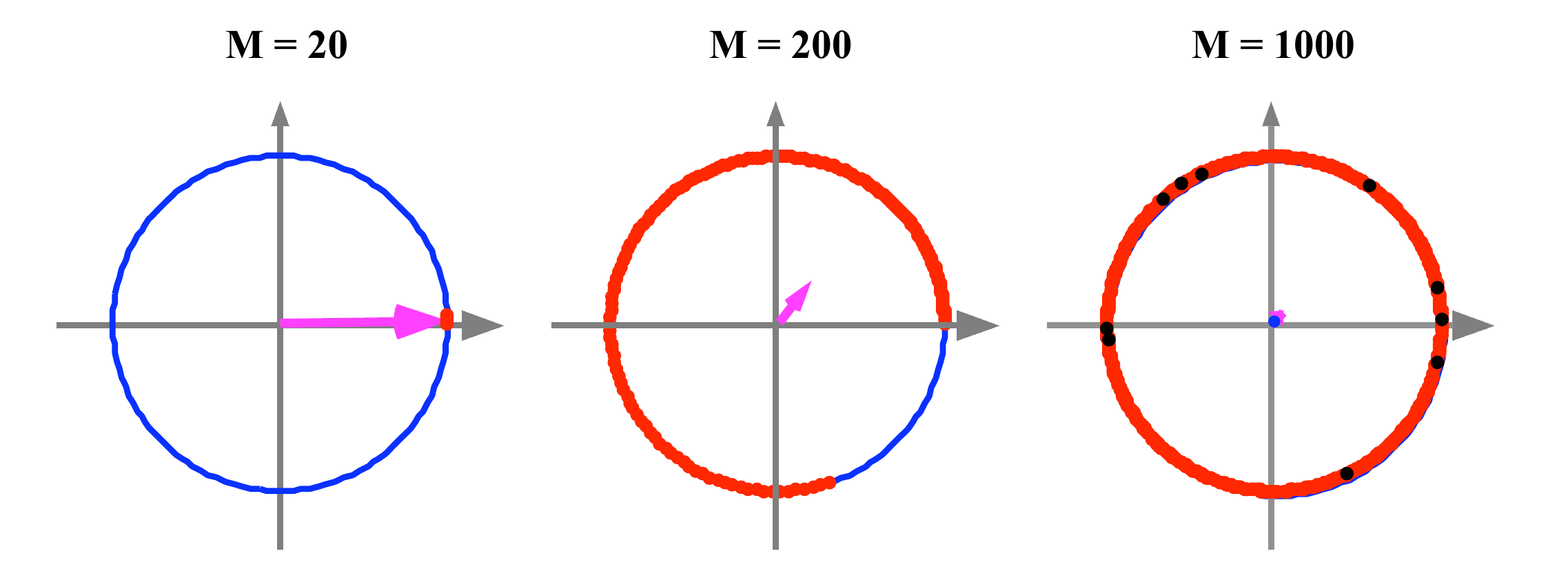}
\end{centering}
\caption{(color online).  Distribution of the phases $\phi_m = \pi m^2\epsilon$ for a fractional part 
$\epsilon(N,l) = 4 \times 10^{-5}$ for different truncation parameters $M$: 
$M = 20$, $M=200$ and $M = 1000$. 
The individual terms of the Gau\ss{ }sums are represented by red dots. 
The arrows from the origin $(0,0)$ represent the resulting truncated Gau\ss{ }sums 
of Eq. (\ref{Tru.Gauss}), whose absolute values are, respectively, 
$|s_{M=20}| \approx 1$,$|s_{M=200}| \approx 0.3155$ and $|s_{M=1000}| \approx 0.0770$. 
In (c), the black points on the circle are the random phases created by a randomized procedure:
10 values of $m$ were randomly chosen from the range $ [0, M_{max}]$.
Their sum is $ \approx 0.0023$, very close to the origin.}
\label{M.principle}
\end{figure}

Fig. \ref{M.principle} illustrates how ghost factors occur for small  $\epsilon$:
the three parts of the figure show the distribution of the different terms of a Gau\ss{ } sum
in the complex plane for (a) $M=20$, (b) 200, and (c) 1000 terms for $\epsilon = 4 \times 10^{-5}$:
only in the third case are the phases of the terms sufficiently well distributed that the sum
(shown as the blue line) approaches zero.

% \section{Randomized procedure}

\emph{Randomized procedure ---} 
Suppressing all ghost factors of truncated Gau\ss{ }sums below the threshold $1/\sqrt{2}$, 
requires $M \sim  \sqrt[4]{N}$ \cite{M-Stefanak:370aa}. 
This may still be too large for experimental factorization of large integers, 
e.g., $M \approx 1000$ for the factorization of a 12-digit integer requires
extreme precision in the experimental implementation and may lead to excessive decoherence.
However, the number of terms in the sum can be reduced significantly
below the $\sqrt[4]{N}$ threshold by a suitable choice of the terms that are evaluated:
instead of evaluating all terms with $0 \le m \le M$, we randomly choose a relatively small fraction
of the terms with $m \in  [0, M  \sim \sqrt[4]{N}]$.

As a test of this procedure, we select a ``difficult" case,
where the conventional procedure requires a large truncation parameter.
Such cases occur, e.g., for numbers that are products of neighboring primes.
We chose as an experimental example the product of the 100000$^{\mathrm{th}}$ 
and 100001$^{\mathrm{th}}$ prime, 
$$
N = 1689259081189 = 1299709 \times 1299721.
$$

In the experiments, we implement the procedure by adding nuclear spin coherence,
using the $^1$H ($I=\frac{1}{2}$) nuclear spins of water, diluted in D$_2$O. 
The nuclear system was contolled by suitable radio-frequency (rf) magnetic fields. In the rotating frame, 
an rf pulse with duration $\tau$, amplitude $\omega$ and phase $\phi_m$ generates the unitary operator
\begin{equation}
U_m = \exp(-i \theta (I_x \cos \phi_m + I_y \sin \phi_m)) .
\end{equation}
The lower index $m$ indicates that the Hamiltonian is specific for each term in the  Gau\ss{ }sum.
The flip angle $\theta = \omega \tau$ represents the absolute value and $\phi_m$ the
phase of a complex number in the series, corresponding to 
$\phi_m(l) = 2 \pi m^2\frac{N}{l}$ in the Gau\ss{ } sum. The sum was realized in the experiment by applying a sequence of $M+1$ rf pulses with small flip-angle 
to the spins, with the phase of each pulse equal to that of the corresponding term of the Gau\ss{ }sum.
A short delay ($5 \mu$s) was inserted between the pulses.
The combined effect of these pulses can be described by the propagator $U(l)  =  U_M \cdots U_0$.

In the limit of small flip angles, $M \theta \ll 1$, 
the operators in the exponent approximately commute and the propagator can be approximated by 
\begin{eqnarray}
U(l) \approx \exp \{-i \theta  \sum_{m=0}^M [I_x \cos\phi_m(l) + I_y \sin\phi_m(l)]\} .
\end{eqnarray} 
If $l$ is a factor of $N$, then $\phi_m(l) = 2k \pi$ with $k$  integer and all $M+1$ pulses 
have the same phase ($\phi = 0$).
In this case, the combined effect of the pulses is 
$U(l) \approx e^{-i \theta (M+1) I_x } $.
If it is applied to the thermal equilibrium state, it creates transverse magnetization $I_y$,
with an amplitude $\propto \theta (M+1)$.
If $l$ is not a factor of $N$, the individual signals interfere destructively
and the resulting transverse magnetization is close to zero. 
For each experiment, the transverse magnetization generated was recorded as a
free induction decay (FID).

\begin{figure}[tbh]
\begin{centering}
\includegraphics[width=0.99\columnwidth]{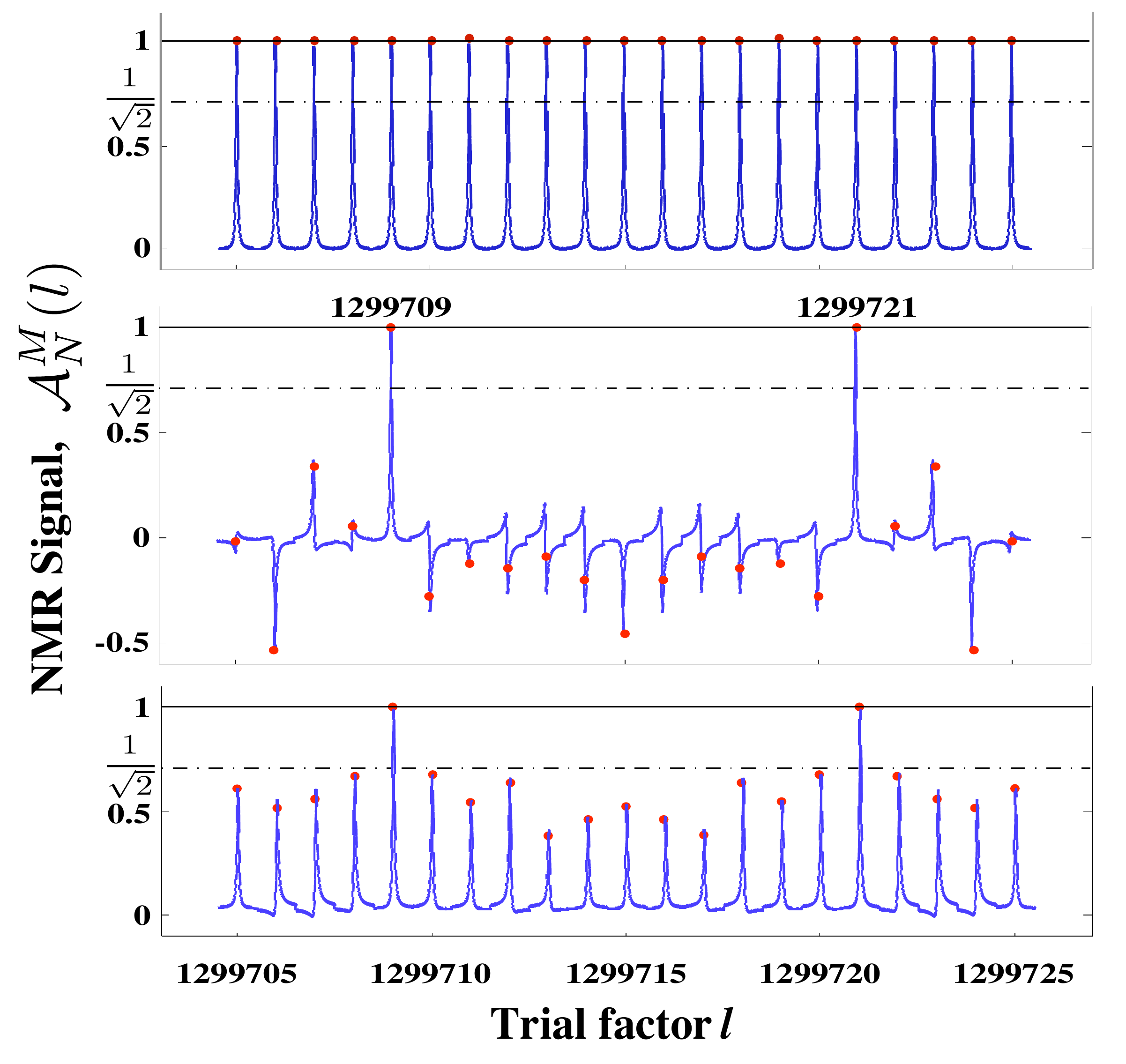}
\end{centering}
\caption{(color online).  Factorization of $N = 1689259081189$ with $\epsilon_{min} \approx 1.693 \times 10^{-5}$.
The upper trace shows the standard truncated Gau\ss{ }sum with the truncation parameter $M=19$,
the middle trace shows the result of the Monte Carlo procedure with 10 randomly chosen  values of $m$  from the range $[0, M_{max} = 1000]$, and the lower trace shows the $5^{\mathrm{th}}$-order truncated  exponential sum $\mathcal{A}5_N^M$ with the truncation parameter $M = 10$. 
While all trial factors masked as true factors in the standard truncated procedure (upper trace),
the true factors are easily found in the Monte Carlo and exponential sum procedure.
} 
\label{exp.large}
\end{figure}

The experiments were carried out on a 500 MHz Bruker Avance II+ NMR spectrometer. Using the standard truncated Gau\ss{ }sum  $\mathcal{A}_N^M(l)$ of Eq. (\ref{Tru.Gauss}) with $M=19$, 
we obtained experimental results that were indistinguishable from the maximal value of 1
for all trial factors close to the real factors [see upper trace in Fig. \ref{exp.large}].
However, if we use the randomly selected $m$-values, as few as 10 terms
are sufficient to suppress all the non-factors well below the threshold of 
$1/\sqrt{2}$, as shown in Fig. \ref{exp.large} (blue spectra and red dots),
while the real factors 1299709 and 1299721 always yielded values close to 1.

\begin{figure}[tbh]
\begin{centering}
\includegraphics[width=0.99\columnwidth]{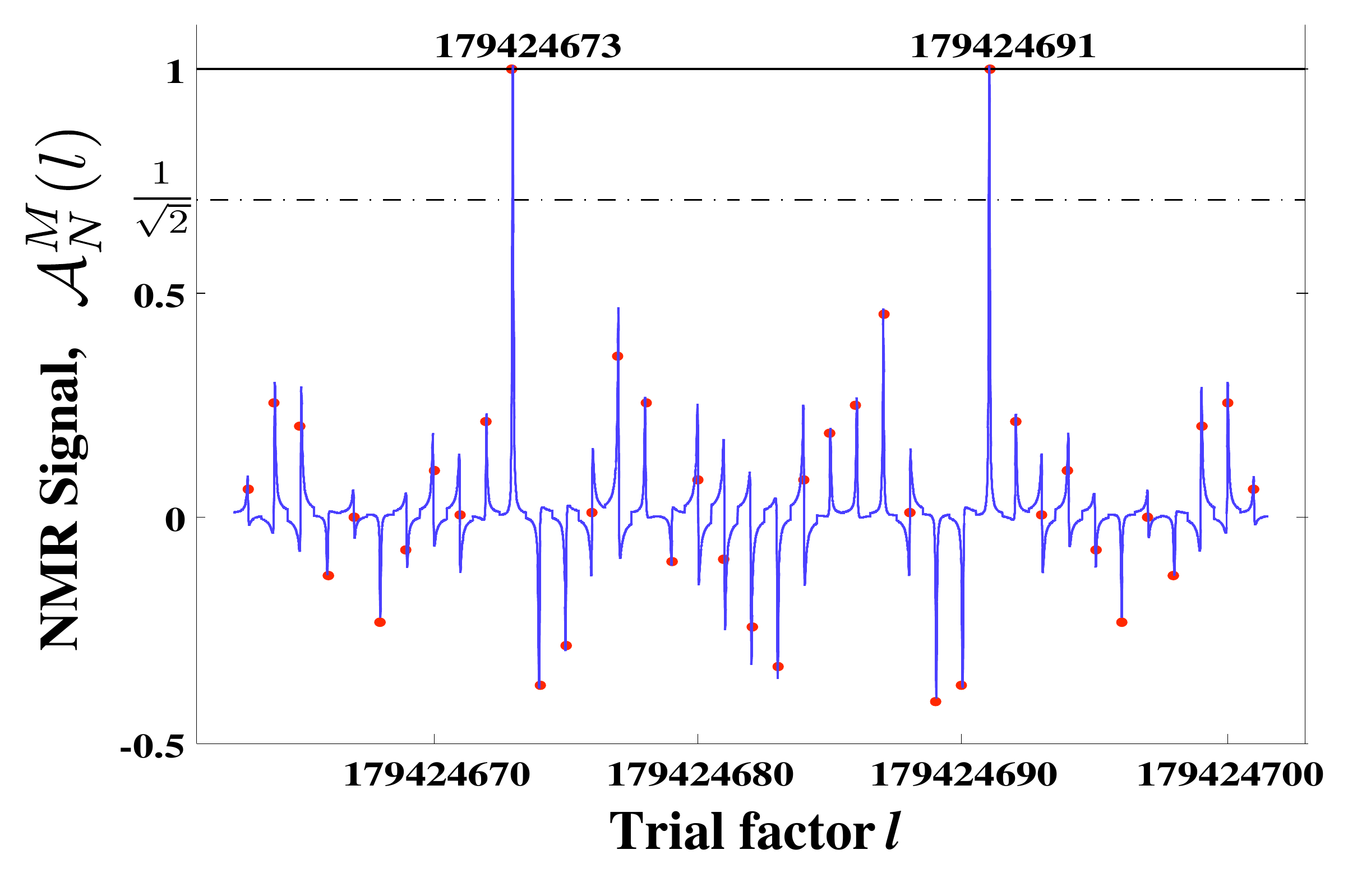}
\end{centering}
\caption{(color online).  Factorization of $N = 32193216510801043$ by the randomized procedure 
with 10 random phases in $[0, M_{max} = 5000]$.} 
\label{exp.large2}
\end{figure}

As the second example, we chose to factorize a 17-digit integer 
$$
N = 32193216510801043 =179424673 \times 179424691.
$$
Randomly choosing 10 values of $m$ from the range $[0,M_{max}=5000]$,
we experimentally evaluated the partial Gau\ss{ }sums for the trial factors $l$ 
between $179424663$ and $179424701$. 
The results, shown in Fig. \ref{exp.large2}, clearly show that the factors 
179424673 and 129424691 are found and no ghost factors appear. 

% \section{Higher-order truncated Gau\ss{ }sum}

\begin{figure}[tbh]
\begin{centering}
\includegraphics[width=0.9\columnwidth]{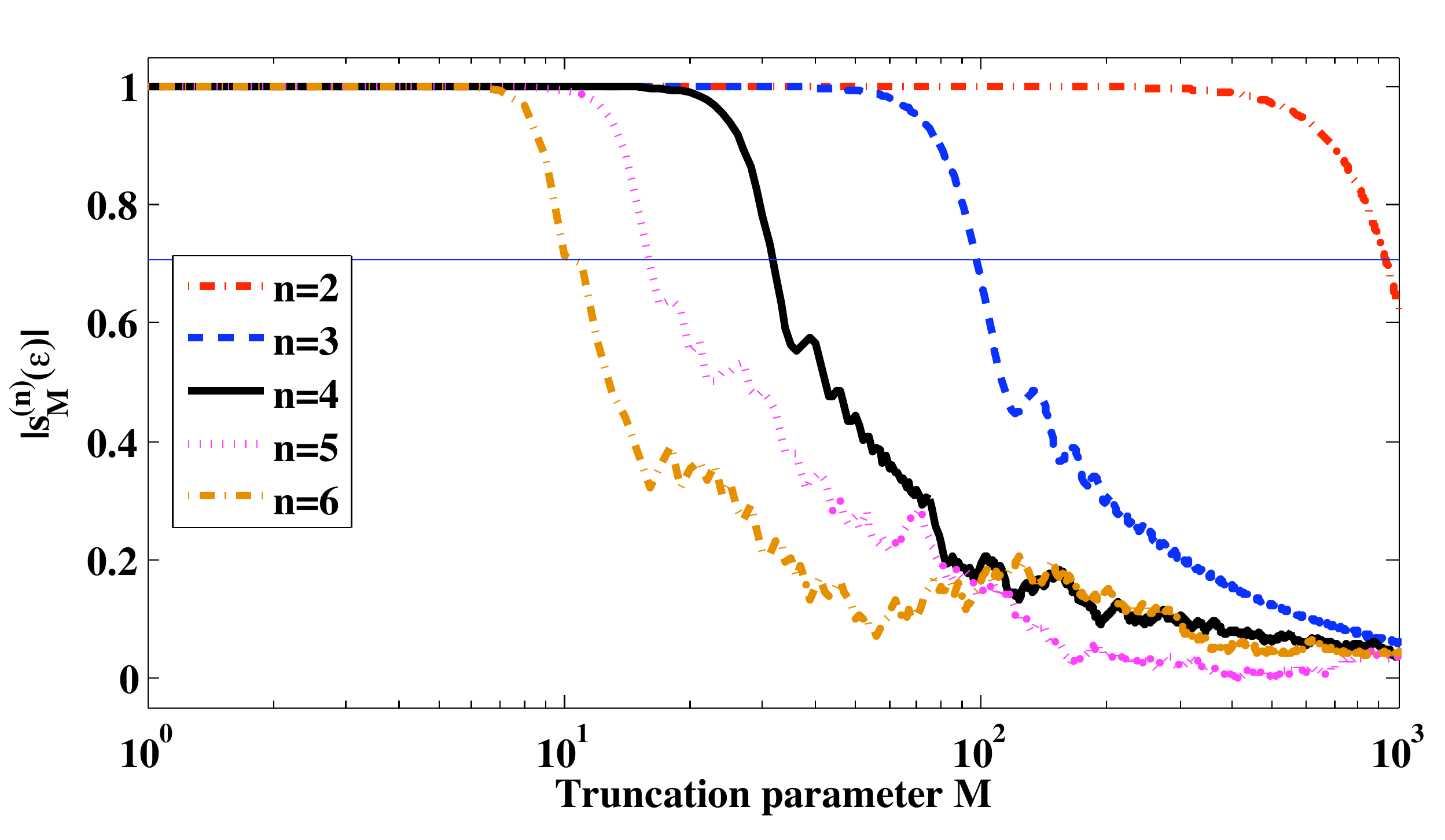}
\end{centering}
\caption{(color online).  Comparison the different orders of Gau\ss{ }sums as a function of the truncation parameter $M$
for a fractional part $\epsilon =1.0 \times 10^{-6}$.} 
\label{mpower}
\end{figure}

\emph{Exponential sums ---} Exponential sums can be defined as
\begin{equation}
\mathcal{A}n_N^M(l) = \frac{1}{M+1}\sum_{m=0}^M\exp\bigg[2\pi i m^n\frac{N}{l}\bigg], n\ge 3.
\label{Tru.Gaussn}
\end{equation}
In terms of the fractional part $\epsilon(N,L)$ of $2N/l$, they are
\begin{equation}
\mathcal{A}n_N^M(l) = s^{(n)}_M(\epsilon) \equiv \frac{1}{M+1}\sum_{m=0}^M\exp\bigg[\pi i m^n \epsilon \bigg], n\ge 3.
\label{sn_M}
\end{equation}
The standard case is recovered for $n=2$.
These higher-order exponential sums can be used for factorization exactly as the second-order function:
again, the factors generate constant phases for all terms and thus the maximum of the sum,
while non-factors ideally generate sums much smaller than unity.
These higher-order exponential sums can provide higher contrast between factors and non-factors, 
even for small truncation parameters $M$. 
As shown in Fig. \ref{mpower}, the higher the order $n$, the smaller the upper bound of $M$
required to suppress the value of $|s_M^{(n)}|$ below the threshold.  

Numerical analysis show that the truncation parameter $M$ required to push all non-factors 
below the threshold $1/\sqrt{2}$ scales with the order $n$ of the exponential sum and the size of
the number $N$ to be factorized as $M \sim \sqrt[2n]{N}$ \cite{Stefanak:2008aa}.
Therefore, to factorize a 12-digit integer, the required value of $M$ decreases from  $10^3$  to 10 
if we use the $6^{\mathrm{th}}$ order function instead of the quadratic truncated Gau\ss{ }sum. 
However the advantage takes the price of the smaller gap between factors and threshold non-factors \cite{Stefanak:2008aa}. The authors also proposed an NMR realization of exponential sums \cite{Stefanak:2008ab}.

We experimentally tested the performance of the higher-order truncated exponential sums $\mathcal{A}n_N^M$,
using the same procedure as for the $n=2$ case.
The lower trace of Fig. \ref{exp.large} clearly shows that this procedure provides an excellent contrast between factors and non-factors,
even for a relatively small number of terms $M=10$.

% \section{Conclusion}

\emph{Conclusion ---} Gau\ss{ }sums \cite{S:1970aa,H:2007aa,H:1980aa} are ubiquitous in number theory and found many applications, such as Plancherel's theorem on finite groups \cite{Yosida:1968aa}, the Talbot effect of classical optics \cite{F:1836aa}, fractional revivals \cite{Leichtle-C:1996ab,Leichtle-C:1996aa}, quantum carpets \cite{Berry-M-V:2001aa} and Josephson junctions \cite{Oppenlander-J:2000aa}. 
Recently, Gau\ss{ }sums were also used for factorization, which is related to the proposal of Clauser and Dowling \cite{Clauser:1996aa} to factor an integer using a familiar YoungÕs N-slit classical interferometer.

In this paper, we presented an experimental investigation 
on the Gau\ss{ }sum factorization scheme for large numbers, 
where ghost factors often appear when the truncation parameter $M$ is relatively small 
(e.g., $M \sim 15 - 20$). 
In these cases, the truncation parameter $M$ must be increased to relatively large numbers,
which is undesirable for experimental implementations.
To circumvent this increase in the required number of terms,
we have introduced a Monte Carlo procedure, where the required number of terms
remains roughly constant, and have used higher-order truncated Gau\ss{ }sums,
whose scaling behavior is much more benign than for the second order function.
While we have used a nuclear spin system for the experimental implementation,
it should be straightforward to apply this scheme to other (quantum or classical) systems.

\textit{Acknowledgement}: 
We acknowledge many useful comments from Wolfgang Schleich.
This work is supported by  the DFG through Su 192/19-1. 

\textit{Note added}: During the preparation of this paper, we became aware of closely related work \cite{Rasel:2008aa,Chatel:2008aa}.

\end{document}